\documentclass[conference]{IEEEtran}

\usepackage[utf8]{inputenc}
\usepackage[T1]{fontenc}
\usepackage{cite}
\usepackage{amsmath,amssymb,amsfonts}
\usepackage{graphicx}
\usepackage{xcolor}
\usepackage{booktabs}
\usepackage{enumitem}
\usepackage[hyphens]{url}
\usepackage{microtype}
\usepackage{multirow}
\usepackage{listings}
\usepackage{dsfont}
\usepackage{algorithm}
\usepackage{algpseudocode}
\algtext*{EndIf}
\algtext*{EndWhile}
\usepackage{xspace}
\usepackage[caption=false,font=footnotesize]{subfig}
\usepackage{balance}
\usepackage[hidelinks]{hyperref}
\usepackage{makecell}
\usepackage[most]{tcolorbox}

\def\BibTeX{{\rm B\kern-.05em{\sc i\kern-.025em b}\kern-.08em
    T\kern-.1667em\lower.7ex\hbox{E}\kern-.125emX}}

\definecolor{mygreen}{rgb}{0,0.6,0}
\definecolor{mygray}{rgb}{0.95,0.95,0.95}
\definecolor{mymauve}{rgb}{0.58,0,0.82}
\definecolor{myblue}{rgb}{0.2,0.42,0.62}
\definecolor{myblue2}{rgb}{0.12,0.45,0.70}
\definecolor{myred}{rgb}{0.8,0,0}
\definecolor{mypurple}{rgb}{0.8,0.8,1}
\definecolor{myyellow}{rgb}{1,0.9,0.6}

\newcommand{\avgsaved}{25.75\%}
\newcommand{\sys}{\mbox{\textsc{biscepter}}\xspace}

\newtcolorbox{takeawaybox}{
    colback=gray!8,
    colframe=black!20,
    boxrule=0.3pt,
    arc=1mm,
    left=1mm,
    right=1mm,
    top=0.5mm,
    bottom=0.5mm
}

\begin{document}

\title{\sys{}: Probability-Driven Bisection for Large-Scale System Software}


\author{
\IEEEauthorblockN{Mingyan Gao\textsuperscript{*}}
\IEEEauthorblockA{\textit{The University of Hong Kong} \\
Hong Kong\\
mingyan.gao@connect.hku.hk}
\and

\IEEEauthorblockN{Celine Wüst\textsuperscript{*}}
\IEEEauthorblockA{\textit{ETH Z\"urich} \\
Switzerland\\
celine.wuest0@gmail.com
}
\and
\IEEEauthorblockN{Zuming Jiang}
\IEEEauthorblockA{\textit{The University of Hong Kong} \\
Hong Kong\\jzuming@hku.hk}
\and
\IEEEauthorblockN{Zhendong Su}
\IEEEauthorblockA{\textit{ETH Z\"urich} \\
Switzerland\\
zhendong.su@inf.ethz.ch}
}

\maketitle
\begingroup
\renewcommand{\thefootnote}{\fnsymbol{footnote}}
\footnotetext[1]{Mingyan Gao and Celine Wüst contributed equally to this work.}
\endgroup
\begin{abstract}
Identifying the bug-inducing commit (BIC) is a fundamental step in regression debugging and a key input to emerging BIC-aware fault-localization pipelines. In practice, BICs are commonly obtained with bisection. Standard bisection selects the median commit of the remaining good--bad interval, thereby balancing commit count. This strategy is optimal under the assumption that each commit is equally likely to be the BIC. 

This paper shows that this assumption does not match real-world BIC histories. We construct a dataset of 8,172 bug reports from GCC, the Linux kernel, and MariaDB. We find a strong temporal skew: across the studied systems, 50\% of BICs lie within the most recent 0.69\% of the report-time commit history. Motivated by this observation, we introduce \sys, a probability-driven bisection approach that uses historical BIC latency as a lightweight prior. Instead of selecting pivots that split the number of remaining commits, \sys selects weighted-median pivots that split estimated BIC probability mass, while preserving the same good--bad oracle and interface as standard bisection. We evaluate \sys on three large-scale systems. The evaluation results show that \sys reduces bisection iterations by 25.75\% on average (up to 55.55\%) compared with standard median bisection, while improving over the baseline in 91.26\% of test cases. Robustness experiments further demonstrate that the benefit remains stable under noisy historical data. 
We expect that our research can effectively save effort in debugging software in practice and, more broadly, benefit future software engineering research by bringing insights about BIC distribution.
\end{abstract}

\section{Introduction}

Regression debugging often begins by localizing a fault in time. When a system exhibits a failure, developers need to determine which historical commit first exhibits failing behavior. The \emph{bug-inducing commit} (BIC) captures this transition point in the commit history~\cite{sliwerski2005changes}. Once identified, a BIC provides a compact and actionable debugging target, enabling developers to inspect a specific change, perform root-cause analysis, or directly revert the offending commit~\cite{wen2020boosting,7816476}. As a result, BIC identification has become a common practice in regression-debugging workflows~\cite{yang2025using}.

Recent fault-localization work further shows that BICs are not only useful for manual debugging, but also provide effective signals for automated localization. BASIC treats the files modified in an identified BIC as fault candidates and shows that this developer-aligned strategy can outperform spectrum-based compiler fault-localization techniques~\cite{yang2025using}. DualTrack further combines BIC-derived historical information with dynamic execution spectra, demonstrating that these two sources capture complementary fault signals and can improve localization accuracy in complex systems such as GCC and LLVM~\cite{li2026dualtrack}. Thus, efficient BIC identification matters both for established regression-debugging workflows and for emerging BIC-aware fault-localization methods.

In practice, BICs are commonly obtained through bisection, as implemented by tools such as \texttt{git bisect}\cite{git-bisect-manual}. 
Given a bug-triggering test, a good commit triggering no bug, and a bad commit triggering the bug, bisection repeatedly tests an intermediate commit to shrink the interval between the good and bad commits until finding the oldest commit where the bug manifests. This simple interface makes bisection broadly applicable: it requires only an ordered commit history and an executable oracle that determines whether the target bug is present. However, each bisection query can be expensive in large-scale systems. Testing a commit requires restoring a compatible build environment, rebuilding the system, and running the resource-intensive test~\cite{saha2017selective}. Reducing the number of tested commits therefore directly reduces the practical cost of BIC identification.

Standard bisection controls search cost by repeatedly testing the median commit of the remaining interval. By halving the number of candidate commits at each step, this strategy minimizes the worst-case search depth when no additional information about the BIC location is available. Implicitly, this approach treats every commit as equally likely to be the bug-inducing commit. Under this assumption, balancing the number of commits on both sides of the pivot is equivalent to balancing the likelihood that the BIC lies on either side. However, if real BICs are not uniformly distributed throughout project commit histories, this equivalence no longer holds. A median pivot may eliminate half of the remaining commits while leaving most of the actual BIC likelihood concentrated on one side of the interval.

\noindent
\underline{\textbf{Empirical Study:} \textit{BIC Distribution.}}
To examine whether this assumption matches practice, we construct a dataset of 8,172 real-world bug reports from three large open-source systems: \emph{GCC}, the \emph{Linux kernel}, and \emph{MariaDB}. 
These systems belong to different software categories (\emph{i.e.}, compiler, OS kernel, and database) but share a strong and consistent BIC temporal skew: 50\% of bugs were introduced by only the most recent 0.69\% of commits in the report-time history. Under a uniform prior, such a narrow recent region should contain about 0.69\% of BICs, not roughly half of them.  This empirical mismatch suggests that median bisection is often balanced by commit count but highly imbalanced by the actual distribution of bug-inducing commits. As a result, early median pivots can spend effort testing commits that are unlikely to be the inducing commit.

\noindent
\underline{\textbf{Core Idea:} \textit{Probability-Driven Bisection.}}
This empirical mismatch motivates \sys, a probability-driven bisection framework. Instead of selecting pivots that split the number of remaining commits, \sys selects pivots that split the estimated BIC probability mass. The key signal is historical BIC latency: how far, in commit-index distance, a bug-inducing commit lies behind the latest commit available when the bug was reported. These historical latencies provide a lightweight project-specific prior over where future BICs are likely to appear. Given this prior, \sys chooses a weighted-median pivot that approximately balances the estimated probability mass of the two possible bisection outcomes. This design targets a different point in the debugging pipeline from techniques such as Fonte~\cite{an2023fonte}, DiWi~\cite{chen2019diwi}, RecBi~\cite{chen2020recbi}, LLM4CBI~\cite{tu2024llm4cbi}, and ODFL~\cite{yang2025odfl}, which require coverage spectra, generated witnesses, test mutation, configuration variation, or language-specific program analysis. \sys preserves the standard bisection workflow, changes only the pivot-selection rule, and uses the same bug-triggering test as the oracle.

We implement \sys as a standalone, language-agnostic bisection tool. We evaluate \sys in a deployment-oriented setting. For each project, we sort BICs chronologically by bug-report time: older records initialize the historical latency prior, a validation partition selects the cutoff that determines when \sys falls back to standard bisection, and the newest records are held out for performance evaluation. The evaluation results on the three large-scale systems show that \sys reduces bisection iterations by an average of \avgsaved{}, up to 55.55\%, compared with standard bisection. It achieves improvement in 91.26\% of test cases. Robustness experiments further demonstrate that \sys remains stable under noisy, missing, and duplicated historical BIC records. These results show that a small change that exploits historical BIC regularities in the pivot-selection rule can significantly improve bisection efficiency while retaining the simplicity and generality of conventional bisection.

This paper makes the following contributions:

\begin{itemize}
\item \textbf{An empirical study of BIC temporal distributions.}
We construct a dataset of 8,172 real-world bug reports from \emph{GCC}, the \emph{Linux kernel}, and \emph{MariaDB}. Across the studied systems,  50\% of bugs are introduced by only the most recent 0.69\% of commits, revealing a strong temporal skew that challenges the equal-likelihood assumption implicit in standard bisection.

\item \textbf{A lightweight probability-driven bisection technique.}
We introduce \sys, which uses historical BIC latencies as a project-specific prior and selects weighted-median pivots that approximately balance estimated BIC probability mass rather than commit count. The approach changes only the pivot-selection rule and preserves the testing interface of standard bisection.

\item \textbf{An evaluation of effectiveness and robustness.}
Across three large-scale systems, \sys reduces bisection iterations by an average of \avgsaved{} (up to 55.55\%) while outperforming standard bisection in 91.26\% of cases. Additional experiments demonstrate that the improvements remain stable under noisy, incomplete, and duplicated historical BIC data.

\end{itemize}

\section{Problem Formulation}
\label{sec:problem}

\subsection{Bisection as Ordered Search}

Given a bug-triggering test, identifying a bug-inducing commit (BIC) can be formulated as an ordered search problem. Let $C = \langle c_0, c_1, \ldots, c_n \rangle$ denote a linearized sequence of commits ordered from older to newer. The first commit $c_0$ is known to be \emph{good}, meaning that the failure is absent, and the last commit $c_n$ is known to be \emph{bad}, meaning that the failure is present. Under the standard monotonicity assumption of bisection, there exists a transition point $c_k \in C$ such that all commits before $c_k$ are good, while $c_k$ and all subsequent commits are bad. The objective is to identify this oldest bad commit $c_k$ using as few test executions as possible.

A bisection strategy maintains a current interval $(g,b]$, where $g$ is the latest tested \emph{good} commit and $b$ is the earliest tested \emph{bad} commit. The BIC is guaranteed to lie in this half-open interval. At each step, the strategy selects a pivot commit $c \in (g,b]$ and runs the bug-triggering test at $c$. If the test passes, then $c$ is good and all commits up to $c$ can be discarded, so the interval becomes $(c,b]$. If the test fails, then $c$ is bad and the BIC must be $c$ or an older commit, so the interval becomes $(g,c]$. The search terminates when the interval contains a single candidate, the oldest bad commit.

This formulation follows the standard assumptions of bisection. The bug-triggering test is assumed to be monotonic over the commit sequence: commits before the BIC are classified as good, while the BIC and later commits are classified as bad. Commits that cannot be evaluated, such as unbuildable or invalid commits, are excluded from the commit sequence or skipped during bisection, as in conventional bisection workflows.  \sys does not change this oracle model: it changes only which pivot commit is selected next.

\begin{figure}[t]
\centering
\includegraphics[width=0.99\linewidth]{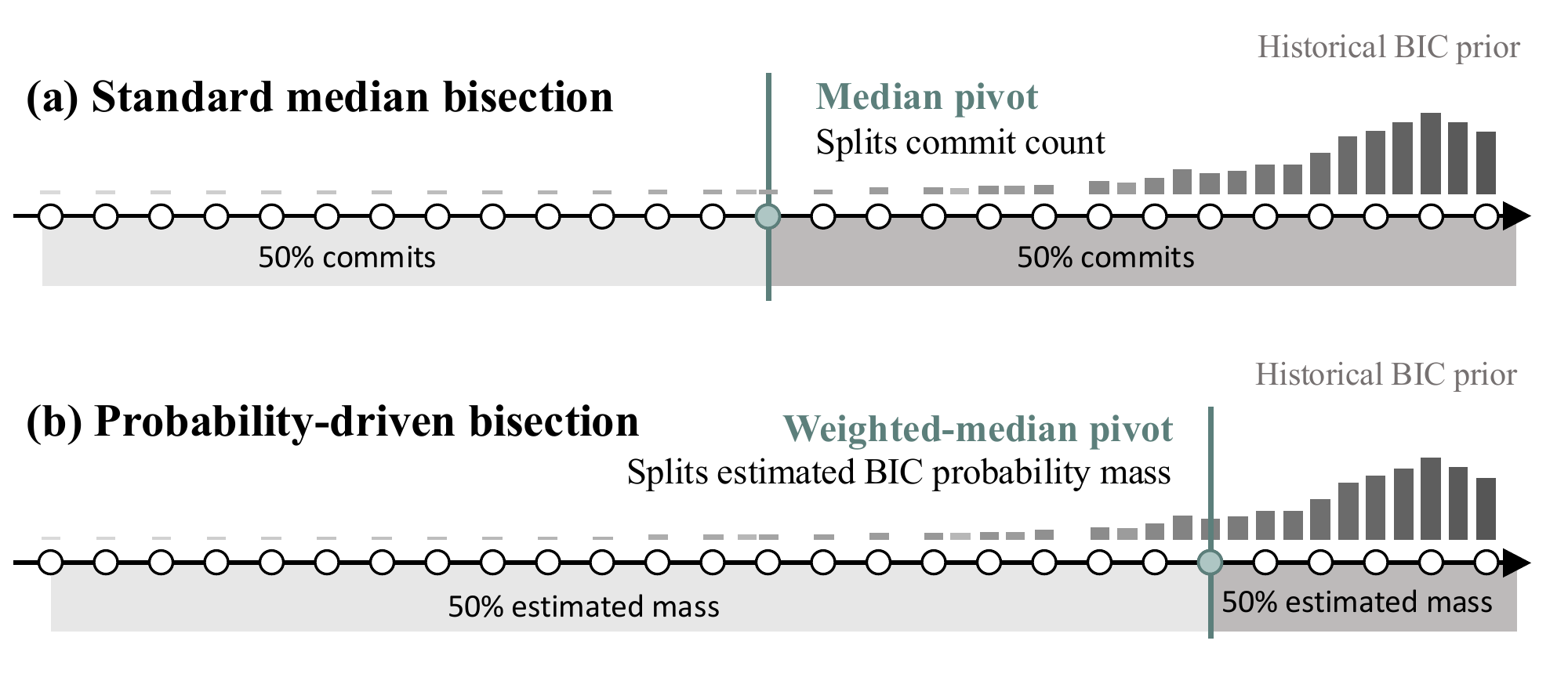}
\vspace{-25pt}
\caption{\label{fig:motivation-bisection} Commit-count partitioning versus probability-mass partitioning.}

\end{figure}

\subsection{From Commit-Count Balance to Probability-Mass Balance}

Standard tools such as \texttt{git bisect}\cite{git-bisect-manual} select the median commit of the remaining interval. This rule is natural when no information is available about where the BIC is likely to occur: each query halves the number of remaining candidate commits and yields an $O(\log n)$ search procedure\cite{courtiel2024theoretical}. In other words, standard bisection optimizes \emph{commit-count balance}, as illustrated in Figure~\ref{fig:motivation-bisection}.

This objective is appropriate when every remaining commit is treated as equally likely to be the BIC. In that case, balancing the number of commits on both sides of the pivot also balances the estimated likelihood that the BIC lies on either side. However, if BICs are more likely to occur in some regions of the history than others, this equivalence no longer holds. A pivot can split the interval evenly by commit count while still leaving most of the BIC likelihood on one side.

Figure~\ref{fig:motivation-bisection} illustrates this mismatch. The standard median pivot divides the candidate commits into two similarly sized sets, but it ignores the fact that the estimated BIC likelihood may be concentrated near recent commits. A probability-aware pivot should instead split the estimated BIC likelihood across the two possible bisection outcomes. The rest of this paper studies whether real BIC histories exhibit such non-uniform structure and how that structure can be used to guide pivot selection.

\section{Empirical Study: The Distribution of Bug-Inducing Commits}
\label{sec_empirical_study}

Section~\ref{sec:problem} shows that median bisection is well aligned with BIC search only when balancing commit count is a reasonable proxy for balancing BIC likelihood. This section tests that assumption empirically. We study whether real-world BICs are spread uniformly over project histories, or whether they concentrate in particular temporal regions that can guide pivot selection. 

\subsection{Dataset Construction}

We constructed a dataset of real-world BICs from the
public bug trackers of \emph{GCC}, the \emph{Linux kernel}, and
\emph{MariaDB}, three large open-source systems with long commit
histories and substantial debugging discussions.

Because BICs are typically discussed in free-form comments rather
than standardized metadata, we use an LLM-assisted extraction pipeline
to identify candidate BIC hashes. Specifically, we use DeepSeek's
R1-Distill-Llama-8B model to parse bug-report discussions and return
either a 40-character SHA-1 hash explicitly described as the commit
that introduced the reported bug, or \texttt{No} when the discussion
does not provide sufficient evidence. The prompt instructs the model
to ignore bug-fixing commits, commits used only for testing, commits
that merely reference the bug report, and any candidate whose role is
ambiguous. The extraction prompt is included in our artifact.

We then apply repository-level validation checks, including hash
canonicalization, commit-existence verification, and chronological
consistency checks, to filter out bug-fixing commits, post-report
testing commits, and invalid hashes. Manual inspection of the first
100 extracted records per project yielded no false positives, providing
evidence that the extraction pipeline is reliable for
aggregate distributional analysis. This process yields 8,172 usable
real-world BIC records, summarized in Table~\ref{tbl_projects}.

\begin{table}[t]
    \caption{Statistics of the projects analyzed.}
    \label{tbl_projects}
    \vspace{-5pt}
    \centering
    \footnotesize
    \setlength{\tabcolsep}{4pt}
    \begin{tabular}{@{}lrrrr@{}}
        \toprule
        \textbf{Name} & \textbf{BICs} & \textbf{Commits} & \textbf{log$_2$(Commits)} & \textbf{Creation Year} \\
        \midrule
        GCC & 7,091 & 221,779 & 17.76 & 1987 \\
        Linux Kernel & 541 & 74,418 & 16.18 & 1991 \\
        MariaDB & 540 & 204,113 & 17.64 & 2009 \\
        \bottomrule
    \end{tabular}
    \vspace{-10pt}
\end{table}

\subsection{BIC Latency Measurement}

We formalize the temporal position of each BIC using \emph{commit latency}. For each bug report, we define its \emph{report-time reference commit} $h$ as the latest commit in the selected project-specific history whose timestamp is no later than the bug-report time. For a bug report with report-time reference commit $h$, the latency of a BIC $b$ is the commit-index distance between $b$ and $h$ in the same commit history. Thus, latency is measured in number of commits. A smaller latency means that the BIC is closer to the recent end of the history available at the time of the bug report.

Let $\mathrm{pos}(c)$ denote the index of commit $c$ in the selected project-specific linearization, ordered from newer to older. For a bug report with reference commit $h$ and BIC $b$, we define the raw BIC latency as
\begin{equation}
\ell_h(b) = \mathrm{pos}(b) - \mathrm{pos}(h).
\end{equation} 
We retain only records where  $b$ lies in the report-time commit history rooted at $h$, so $\ell_h(b) \ge 0$ for all retained BICs.

Because the studied projects have different history lengths, raw commit distances are not directly comparable across projects. Let $N_h$ denote the number of commits in the report-time commit history for reference commit $h$. We define the normalized BIC latency as
\begin{equation}
\tilde{\ell}_h(b) = \frac{\ell_h(b)}{N_h}.
\end{equation}
A normalized latency of $1\%$, for example, means that the BIC lies within the most recent $1\%$ of the commit history at the time of the bug report.

The commit history is a deterministic linearization of the repository history. We measure BIC latency over this linearized history, so each commit has a unique position and the distance between a BIC and the report-time reference commit is well defined. In our study, GCC and the Linux kernel use first-parent histories, while MariaDB uses an all-commit topological order, as reconstructed in Section~\ref{sec:implementation}.

\subsection{Temporal Skew in BIC Latency}

\begin{figure*}[t]
\centering
\includegraphics[width=.92\textwidth]{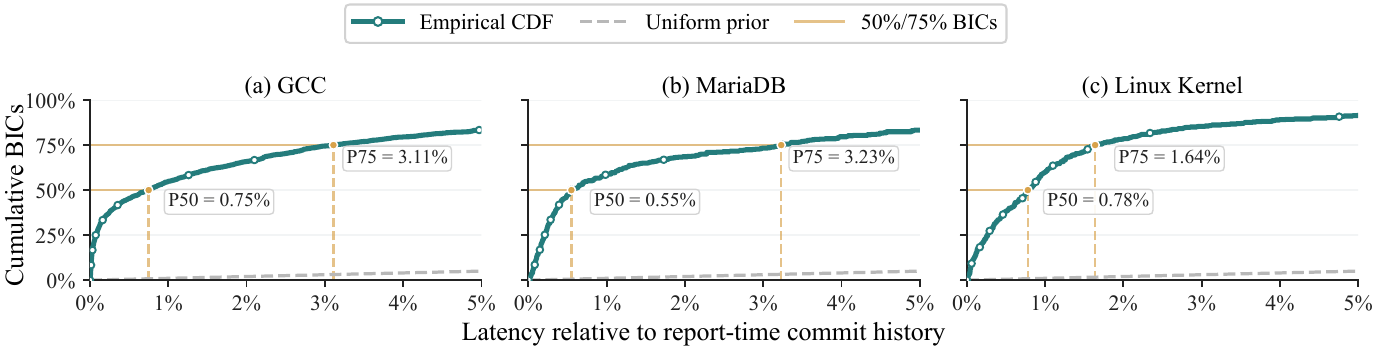}%
\vspace{-3pt}
\caption{Cumulative distributions of normalized BIC latency relative to each report-time commit history.}
\label{fig:commit-latencies-section}
\end{figure*}


Figure~\ref{fig:commit-latencies-section} shows a strong temporal skew in BIC locations. Across the studied projects, the median normalized BIC latency is only 0.69\% on average. That means that 50\% of BICs lie within the most recent 0.69\% of the report-time commit history. The concentration is not limited to the median: even the 75th percentile remains below 4\% for all three projects. 

This pattern is far from what is expected under a uniform prior over commits. If every commit in the commit history were equally likely to be bug-inducing, the most recent 1\% of commits should contain roughly 1\% of BICs. Instead, the observed CDFs place a large fraction of BICs in this narrow recent region. The empirical CDFs therefore rise much faster than the uniform-prior line near the recent end of the history, indicating that commit-count balance is a poor proxy for BIC-likelihood balance in these datasets.

Together, these results show that real-world BICs are consistently concentrated near the recent end of project histories, rather than being spread uniformly over the commit space. This concentration makes historical BIC latency a natural signal for probability-driven pivot selection: if past inducing commits tend to occur at small latencies, future bisection runs can use that pattern to prioritize pivots near regions with higher estimated BIC likelihood.

\section{Approach: Probability-Driven Bisection}
\label{sec_approach}

\begin{figure*}[t]
    \centering
    \includegraphics[width=0.7\linewidth]{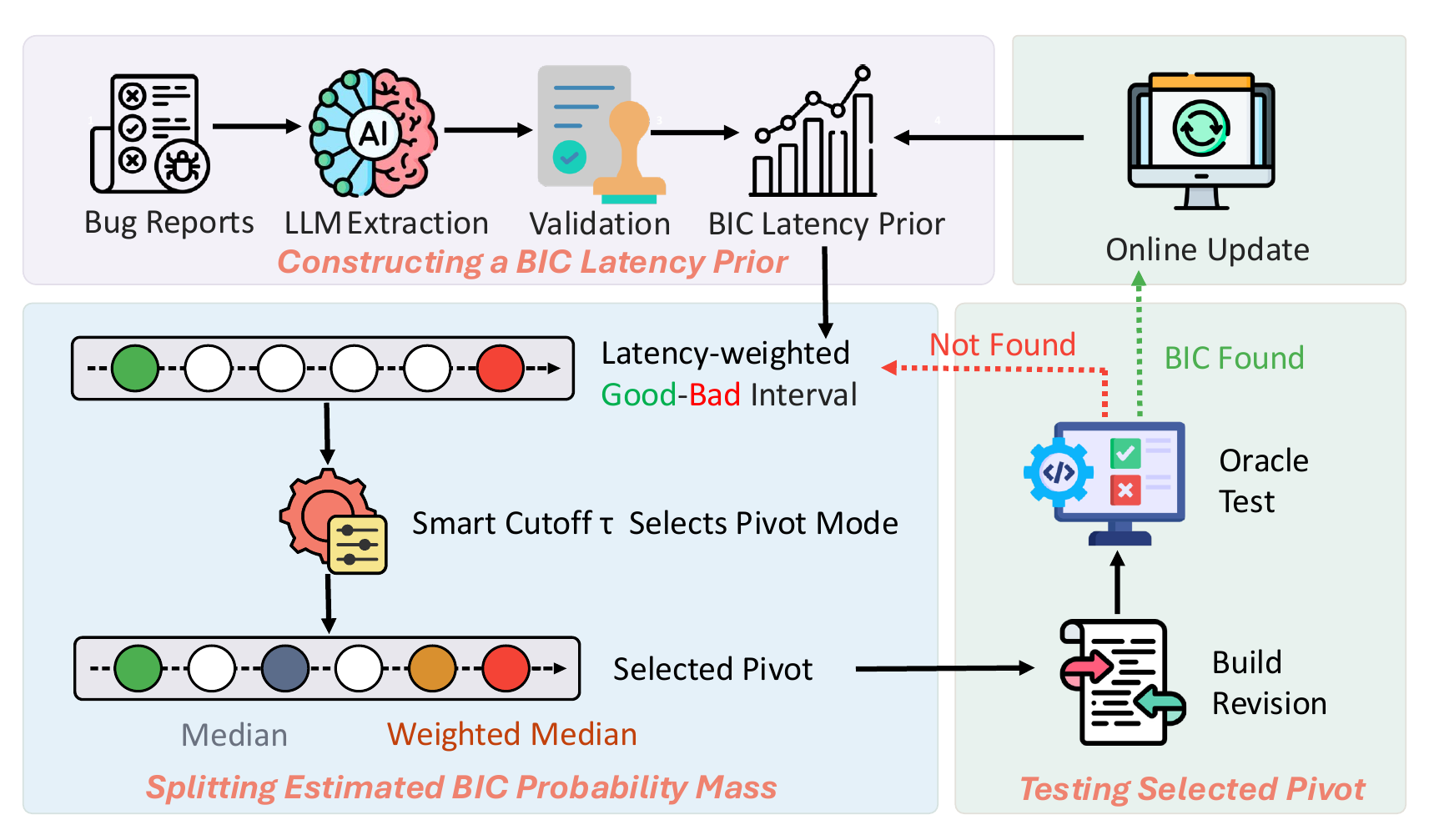}
    \vspace{-3pt}
    \caption{\label{fig:approach-overview} Overview of \sys. Historical BIC latencies define an empirical prior to select probability-balancing pivots. }
\end{figure*}

To exploit the temporal skew observed in Section~\ref{sec_empirical_study}, we propose \sys. \sys retains the ordered-search structure and execution-based oracle of \texttt{git bisect}, but changes the pivot-selection rule. Instead of selecting the commit-count median, \sys estimates where BICs are more likely to occur based on historical BIC latency and selects a pivot that balances estimated BIC probability mass across the two possible bisection outcomes (Figure~\ref{fig:approach-overview}).

\subsection{Estimating Commit Weights from Historical Latency}

Given a bisection interval $I=[g,b]$ and a report-time reference commit $h$, \sys assigns an empirical weight to each candidate commit $c \in I$ based on its latency $\ell(c)$, defined in Section~\ref{sec_empirical_study}. \sys maintains a multiset of known historical BIC latencies $H={\ell_1,\ldots,\ell_n}$ and converts it into an exact-index empirical histogram. Because latency is measured as an integer position in the chosen linearized history, equality in the histogram denotes exact equality of commit-index distance rather than equality of wall-clock time.

The weight of a candidate commit $c$ is the number of known historical BICs with the same latency:
\begin{equation}
    w(c) = \sum_{i=1}^{n} \mathds{1}_{\{BIC[i] = \ell(c)\}} .
\end{equation}

This weighting scheme is intentionally simple. Commits whose latencies have appeared frequently among known BICs receive larger weights, while commits with unseen latencies receive zero weight. Because the weight depends only on commit position in the linearized history, it requires no coverage collection, test mutation, witness generation, or language-specific program analysis.

If no candidate commit in the current interval matches any historical BIC latency, then the interval has no historical support under this exact-index histogram. In that case, \sys falls back to standard median bisection. This fallback prevents the historical prior from making arbitrary pivot choices in regions where the latency data provides no signal. After a bisection run identifies a new BIC, \sys appends its latency to $H$, allowing the empirical distribution to update over time.

\subsection{Selecting Probability-Balancing Pivots}

Consider the current good--bad commit interval $(g,b]$, where $g$ is the
known-good commit and $b$ is the known-bad commit. For a candidate
pivot $c \in (g,b]$, the test oracle has two possible outcomes. If $c$
is classified as \emph{bad}, then the BIC is either $c$ or an older
commit, and the remaining interval becomes $(g,c]$. If $c$ is classified
as \emph{good}, then the BIC must lie in a newer commit, and the
remaining interval becomes $(c,b]$.

We define the estimated BIC mass of any half-open interval $(a,b]$ as
\begin{equation}
\mathsf{mass}(a,b) = \sum_{x \in (a,b]} w(x).
\end{equation}
Under pivot $c$, the two possible outcomes therefore retain estimated
masses $\mathsf{mass}(g,c)$ and $\mathsf{mass}(c,b)$. \sys selects the
pivot that best balances these two retained masses:
\begin{equation}
\label{eq:weighted_median}
c^* = \arg\min_{c \in (g,b]}
\left| \mathsf{mass}(g,c) - \mathsf{mass}(c,b) \right|.
\end{equation}
This pivot is a weighted median of the current interval: it balances
estimated BIC probability mass rather than raw commit count. When
historical BICs are concentrated near recent commits, this rule shifts
early pivots toward high-mass regions that standard bisection
reaches only after several count-balanced splits.

To compute this objective efficiently, we index the current
commit interval from older to newer and represent it by
$(i_g,i_b]$. Let
\begin{equation}
T = \mathsf{mass}(i_g,i_b)
\end{equation}
be the total estimated BIC mass in the current interval. For any
candidate pivot $j \in (i_g,i_b]$, the two retained masses satisfy
\begin{equation}
\mathsf{mass}(i_g,j) + \mathsf{mass}(j,i_b) = T.
\end{equation}
Therefore, minimizing Equation~\ref{eq:weighted_median} is equivalent
to finding a pivot whose left-side mass is closest to $T/2$:
\begin{equation}
\label{eq:prefix_mass_objective}
j^* =
\arg\min_{j \in (i_g,i_b]}
\left|2\mathsf{mass}(i_g,j)-T\right|.
\end{equation}

During bisection, the commit weights are fixed. \sys therefore
precomputes an inclusive prefix-mass array $P$, where $P[k]$ stores the
total weight of all commits with indices at most $k$ in the current
linearized commit history:
\begin{equation}
P[k] = \sum_{x \le k} w(x).
\end{equation}
This array allows \sys to compute the left-side mass of any pivot by
two array lookups and one subtraction:
\begin{equation}
\mathsf{mass}(i_g,j)=P[j]-P[i_g].
\end{equation}
Thus, each interval-mass query takes constant time, independent of the
number of commits in the interval.

Since $\mathsf{mass}(i_g,j)$ is monotone nondecreasing as $j$ moves from
older to newer commits, \sys searches for the optimal commit index specified in Equation~\ref{eq:prefix_mass_objective} by finding the first index $j_0$
whose left-side mass reaches half of the total mass:
\begin{equation}
\label{eq:p_weighted_median}
P[j_0]-P[i_g] \ge T/2.
\end{equation}
Because commit weights are discrete, this first crossing point is not
always the closest balance point. The optimum of
Equation~\ref{eq:prefix_mass_objective} must be either $j_0$ or its
immediate predecessor inside the interval. 
Algorithm~\ref{alg:weighted_median_pivot} shows the procedure of pivot selection in \sys. Specifically, \sys conducts binary search to find the first commit index $j_0$ satisfying Condition~\ref{eq:p_weighted_median}. \sys compares $j_0$ and its immediate predecessor $j_1$, returning the one with the smaller imbalance as the selected pivot for our probability-driven bisection.

\begin{algorithm}[t]
\caption{Selecting a Probability-Balancing Pivot}
\label{alg:weighted_median_pivot}
\small
\begin{algorithmic}[1]
\Require interval indices $(i_g,i_b]$, prefix-mass array $P$
\Ensure pivot index $j^*$
\State $T \gets P[i_b]-P[i_g]$
\If{$T=0$}
    \State \Return $\mathrm{Median}(i_g,i_b)$
\EndIf
\State $l \gets i_g+1$, $r \gets i_b$
\While{$l<r$}
    \State $m \gets \lfloor(l+r)/2\rfloor$
    \If{$P[m]-P[i_g] \ge T/2$}
        \State $r \gets m$
    \Else
        \State $l \gets m+1$
    \EndIf
\EndWhile
\State $j_0 \gets l$
\If{$j_0=i_g+1$}
    \State \Return $j_0$
\EndIf
\State $j_1 \gets j_0-1$
\State $\Delta_0 \gets |(P[j_0]-P[i_g])-T/2|$
\State $\Delta_1 \gets |(P[j_1]-P[i_g])-T/2|$
\If{$\Delta_0 \le \Delta_1$}
    \State $result \gets j_0$
\Else
    \State $result \gets j_1$
\EndIf
\State \Return $result$
\end{algorithmic}
\end{algorithm}

\subsection{Adaptive Fallback: Smart Bisection Cutoff}

Historical latency is most useful in the early stages of bisection, when
the interval is large and coarse temporal patterns can meaningfully
affect pivot selection. As the interval shrinks, however, the exact-index
histogram becomes sparse: only a small number of candidate commits
remain, and local historical weights may become noisy or unavailable.

To avoid overusing a sparse prior, \sys introduces a \emph{smart
bisection cutoff} $\tau$. During the first $\tau$ bisection iterations,
\sys uses the probability-balancing rule in
Algorithm~\ref{alg:weighted_median_pivot}. Once $t \ge \tau$, it switches
to the standard commit-count median. Section~\ref{sec_evaluation} studies
how $\tau$ is selected and how sensitive \sys is to this choice.

\begin{algorithm}[t]
\caption{\sys Probability-Driven Bisection}
\label{alg:biscepter}
\small
\begin{algorithmic}[1]
\Require good–bad interval $(g,b]$, test oracle $\mathrm{Test}$, reference commit $h$, historical latencies $H$, cutoff $\tau$
\Ensure oldest bad commit
\State $w \gets \mathrm{ConstructWeight}(H)$
\State $\mathsf{mass} \gets \mathrm{ConstructMass}(w)$
\State $t \gets 0$
\While{$b-g>1$}
\If{$t < \tau$}
\State $p \gets \mathrm{ProbabilityPivot}(g,b,\mathsf{mass})$
\Else
\State $p \gets \mathrm{Median}(g,b)$
\EndIf
\State $o \gets \mathrm{Test}(p)$

\If{$o=\textsc{Bad}$}
\State $b \gets p$
\Else
\State $g \gets p$
\EndIf
\State $t \gets t+1$
\EndWhile
\State $\ell_{\mathrm{new}} \gets \mathrm{Latency}(b,h)$

\State $H \gets H \cup \{\ell_{\mathrm{new}}\}$
\State \Return $b$
\end{algorithmic}
\end{algorithm}

\subsection{Illustrative Example}
\label{sec_worked_example}

We use \emph{GCC} bug~\#92421 to illustrate how probability-balanced pivots change the search trajectory. Both standard bisection and \sys start from the same report-time good--bad interval, use the same deterministic oracle, and terminate at the same developer-reported BIC; they differ only in which pivot commit is tested next. The initial search interval contains 172{,}573 candidate commits. The developer-reported BIC is \texttt{b8dd4aa1}, whose latency is 181 commits from the report-time reference commit, corresponding to 0.105\% of the commit history. Thus, the inducing commit lies very close to the recent end of the report-time history, where the empirical latency prior from Section~\ref{sec_empirical_study} assigns substantial BIC probability mass.

\begin{figure}[t]
    \centering
    \includegraphics[width=.90\columnwidth]{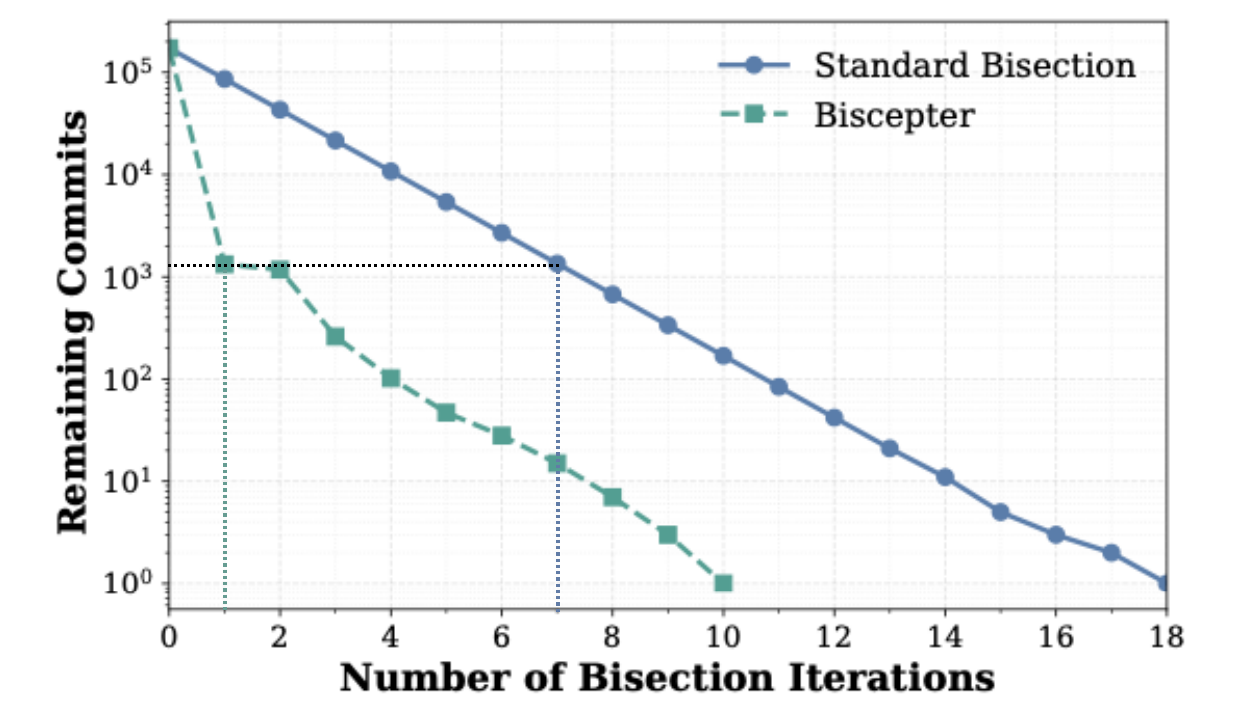}
    \vspace{-3pt}
    \caption{Search-space contraction for \emph{GCC} bug~\#92421. 
    \sys reaches the developer-reported BIC in 10 tests, compared with 18 tests for standard median bisection.}
    \label{fig:example-search-space}
\end{figure}

Figure~\ref{fig:example-search-space} shows the resulting search-space contraction. Standard bisection is balanced by commit count: after one test, it reduces the interval from 172{,}573 to 86{,}286 candidates, and after seven tests it still leaves 1{,}348 candidates. \sys follows the same good--bad update rule, but its early pivots balance historical BIC mass rather than commit count. Because the prior places much of the mass near recent latencies, \sys leaves only 1{,}327 candidates after one test and 15 candidates after seven tests. At this point, the validation-selected cutoff for \emph{GCC} has been reached, so \sys switches back to standard median bisection and finishes the narrowed residual search, identifying the BIC in 10 tests instead of 18.

\section{Implementation}
\label{sec:implementation}

We implement \sys as a standalone evaluation pipeline over deterministic project-specific commit sequences. The pipeline maps each developer-reported BIC into a latency coordinate, simulates standard bisection and \sys over the same good--bad interval, and updates the historical latency ledger chronologically.

The main implementation choice is how to linearize each repository commit history. We choose one deterministic commit sequence per project before evaluation. GCC and the Linux kernel use first-parent histories, which follow the mainline integration path and match the mainline-oriented bisection setting modeled in our evaluation. MariaDB instead uses an all-commit topological order because its history contains large merge commits that integrate entire minor-version development histories. Under a first-parent traversal, many merged development commits are compressed
behind a small number of merge commits and do not appear in the final commit sequence; only 113 MariaDB BIC records map into that sequence. The all-commit topological order preserves these merged commits while still assigning each commit a deterministic position. In all cases, the policy is fixed before evaluation, and standard bisection and \sys use exactly the same sequence within each project.


Our evaluation uses trace-driven simulation rather than executing each candidate commit. For each retained bug, the reconstructed BIC position defines the deterministic bisection outcome for any pivot, allowing us to count how many oracle queries each strategy requires on the same search interval. We therefore report bisection iterations rather than wall-clock debugging time: actual time depends on project-specific build environments, rebuild costs, and test execution time, while iteration count isolates the effect of pivot selection. 
\section{Evaluation}
\label{sec_evaluation}

Our evaluation examines whether probability-guided pivot selection reduces bisection effort. Following the trace-driven simulation
described in Section~\ref{sec:implementation}, we use the number of bisection iterations as the primary cost metric, since each iteration
corresponds to one oracle query on a historical commit. We address three research questions:

\begin{enumerate}[start=1,label={\bfseries RQ\arabic*},font=\itshape, 
leftmargin=*]
    \item \textbf{Effectiveness:} How much does \sys reduce bisection iterations over standard bisection? (Section~\ref{sec_effectiveness})
    \item \textbf{Sensitivity:} How does the smart bisection cutoff affect probability-guided search? (Section~\ref{sec_hyperparam})
    \item \textbf{Robustness:} How resilient is \sys to noisy, missing, or duplicated historical records? (Section~\ref{sec_robustness})

\end{enumerate}

\subsection{Experimental Setup}
\label{sec_eval_setup}

We evaluate \sys in a setting that mirrors how it will be used after deployment: a project first accumulates historical BICs, uses past debugging outcomes to choose a configuration, and then applies the fixed configuration to future faults. To simulate this process without leaking future information, we sort the collected BIC records by bug-report time and split the BICs of each project into three consecutive partitions. The oldest 60\% initializes the historical BIC-latency prior, the next 10\% is used for configuration selection, and the newest 30\% is reserved for final testing.

The only configuration selected from validation data is the smart bisection cutoff $\tau$. For each project, we enumerate $\tau \in {0,1,\ldots,9,\infty}$ on validation bugs and select the value with the lowest average number of bisection iterations. The selected cutoffs are fixed before evaluating the held-out test partition.

The held-out test partition is processed online, matching the intended use of \sys over time. At the start of testing, the latency prior contains only information available before the test period, namely the training records and the validation records used for cutoff selection. Test bugs are then evaluated chronologically. After each test bug is resolved, its true BIC latency is appended to the historical ledger and becomes available for subsequent test bugs. Thus, \sys can learn from newly resolved faults during testing, but no test bug is used to select $\tau$ or any other configuration parameter.

After coordinate reconstruction and filtering, the evaluation contains 7,091 usable \emph{GCC} bugs, 540 \emph{MariaDB} bugs, and 541 \emph{Linux kernel} bugs. The held-out test partitions contain 2,128, 162, and 163 bugs, respectively.

\subsection{RQ1: Comparison with Standard Bisection}
\label{sec_effectiveness}

We use standard median bisection as the controlled baseline for RQ1,
because it has the same input interface as \sys and differs only in
how pivots are selected. Other BIC-identification or fault-localization
techniques, such as FONTE, DiWi, RecBi, LLM4CBI, and ODFL, require
additional fault-specific artifacts, including coverage spectra,
generated witnesses, test mutation, configuration variation, or
language-specific analysis~\cite{an2023fonte,chen2019diwi,chen2020recbi,tu2024llm4cbi,yang2025odfl}.
They therefore evaluate a different design point rather than a
drop-in replacement for standard bisection. This controlled comparison
isolates the effect of replacing commit-count partitioning with
probability-mass partitioning.

\begin{figure*}[t]
\centering
{%
\includegraphics[width=.30\textwidth]{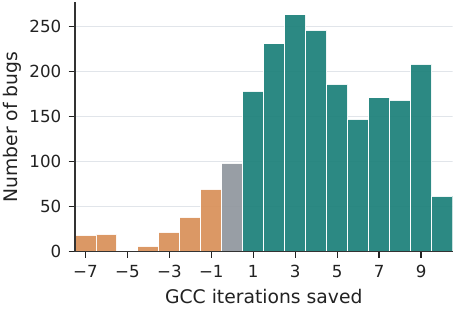}%
}\hfill
{%
\includegraphics[width=.30\textwidth]{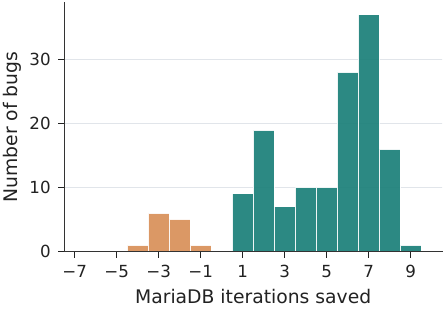}%
}\hfill
{%
\includegraphics[width=.30\textwidth]{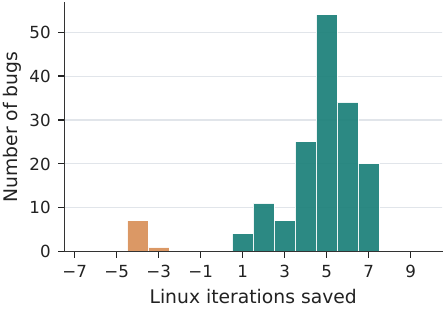}%
}
\vspace{-3pt}
\caption{Difference in bisection iterations between \sys and standard median-based bisection. Positive values indicate iterations saved by \sys.}
\label{fig:experiment-results}
\end{figure*}

\smallskip
\noindent\textbf{Saved Iterations.}
Using the validation-selected cutoff, we evaluate \sys on the held-out
test set. Table~\ref{tbl_results} reports average iterations, absolute and relative savings, the percentage of improved cases, and the maximum per-bug savings. Probability-guided pivot selection consistently reduces bisection effort across all three systems.
Overall, \sys lowers the average number of bisection iterations from
17.12 to 12.73, yielding a 25.75\% reduction over standard median
bisection. The savings are similar in magnitude across projects:
\sys saves 4.12 iterations per bug on \emph{GCC}, 4.59 on
\emph{MariaDB}, and 4.48 on the \emph{Linux kernel}, corresponding to
relative reductions of 23.43\%, 25.88\%, and 27.94\%, respectively.
Figure~\ref{fig:experiment-results} shows that these improvements are
not driven by a few unusually favorable cases. At the per-bug level,
\sys improves over standard bisection in 87\%--95\% of held-out cases,
with the largest savings reaching 10 iterations for \emph{GCC}, 9 for
\emph{MariaDB}, and 7 for the \emph{Linux kernel}.

\begin{table*}[t]
\caption{Held-out test results comparing \sys with standard bisection. }
\label{tbl_results}
\centering
\setlength{\tabcolsep}{6pt}
\renewcommand{\arraystretch}{0.95}
\footnotesize
\begin{tabular}{@{}lrrrrrrrr@{}}
\toprule
\textbf{Project} &
\textbf{N} &
\textbf{Base Iter.} &
\textbf{\sys Iter.} &
\textbf{Saved Iter.} &
\textbf{Rel. Red.} &
\textbf{Imp.(\%)} &
\textbf{Max Saved} &
\textbf{Max Rel. Red.}\\
\midrule
GCC     & 2,128 & 17.60 & 13.48 & 4.12 & 23.43\% & 87.36\% & 10 & 55.55\%\\
MariaDB & 162   & 17.74 & 13.15 & 4.59 & 25.88\% & 91.33\% & 9 & 50.00\%\\
Linux   & 163   & 16.03 & 11.55 & 4.48 & 27.94\% & 95.09\% & 7 & 43.75\%\\
\midrule
\textbf{Avg.} & -- & 17.12 & 12.73 & 4.39 & 25.75\% & 91.26\% & -- & --\\
\bottomrule
\end{tabular}
\end{table*}

\begin{figure*}[t]
\centering
\includegraphics[width=.92\textwidth]{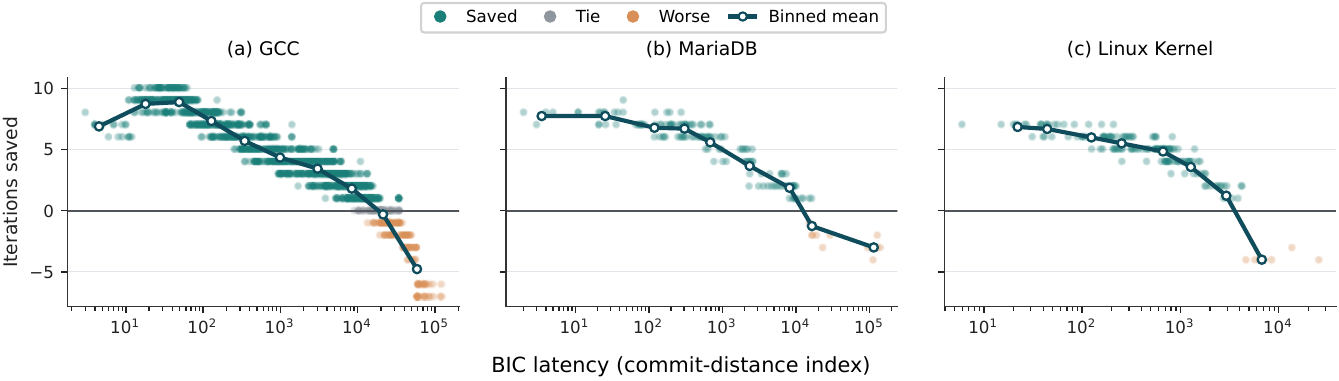}
\vspace{-3pt}
\caption{Relationship between BIC latency and iteration savings on held-out bugs. Each point is a bug; positive values indicate fewer iterations than standard median bisection. The line shows the binned mean over log-scaled BIC latency.}
\label{fig:benefit-latency}
\end{figure*}

\smallskip
\noindent\textbf{Effects of BIC Latency.}
To understand where the savings come from, we relate iteration savings
to the true BIC latency in Figure~\ref{fig:benefit-latency}. \sys achieves
stable positive savings in the recent-latency region, which dominates
the empirical BIC distribution observed in Section~\ref{sec_empirical_study}.
The binned means remain positive across this dense portion of the history,
indicating that the historical prior benefits not only the very newest
commits but also a broader range of recently introduced BICs. This pattern
matches the design of probability-guided bisection: when the true BIC lies
in a latency range that has frequently contained past BICs, weighted-median
pivots can discard low-mass portions of the interval earlier than
commit-count median pivots. As BIC latency grows, the benefit decreases
and penalty cases concentrate in the older tail, where the true BIC lies
far from the high-probability region learned from history.

\smallskip
\noindent\textbf{Penalty Case Study.}
A representative penalty case illustrates this pattern. In \emph{GCC} bug~\#96029, reported in 2020, the true BIC was introduced in 2013. In commit-distance terms, the true BIC
lies at latency 54,978, or 30.90\% of the report-time searchable
history. This location is much older than the region emphasized by the learned GCC prior: before this bug is processed, 3{,}378 of 4{,}645 prior records in the active interval fall within latency 5{,}000, while only 45 fall in the 50{,}001–75{,}000 bucket containing this BIC. Consequently, \sys's first six probability-driven pivots test recent high-mass locations and classify them as bad, proving that the BIC is older than the prior expected. The seventh probability-driven pivot finally crosses the true BIC, but the remaining interval still contains 11{,}476 commits. After the cutoff is reached, standard bisection finishes the residual search, yielding 21 tests compared with 17 under standard bisection. This case shows the main downside risk of \sys: when the true BIC lies in the sparse older tail of the historical distribution, probability-driven pivots can spend several early tests ruling out recent high-mass regions. The smart cutoff bounds this risk by returning to median bisection after a fixed number of probability-driven steps. 

\begin{takeawaybox}
\textbf{Answer to \textit{RQ1}:} \sys reduces iterations by 25.75\% on average across projects (up to 55.55\%) compared with standard bisection.
\end{takeawaybox}

\subsection{RQ2: Sensitivity to the Smart Bisection Cutoff}
\label{sec_hyperparam}

\begin{figure*}[t]
\centering
\includegraphics[width=.92\textwidth]{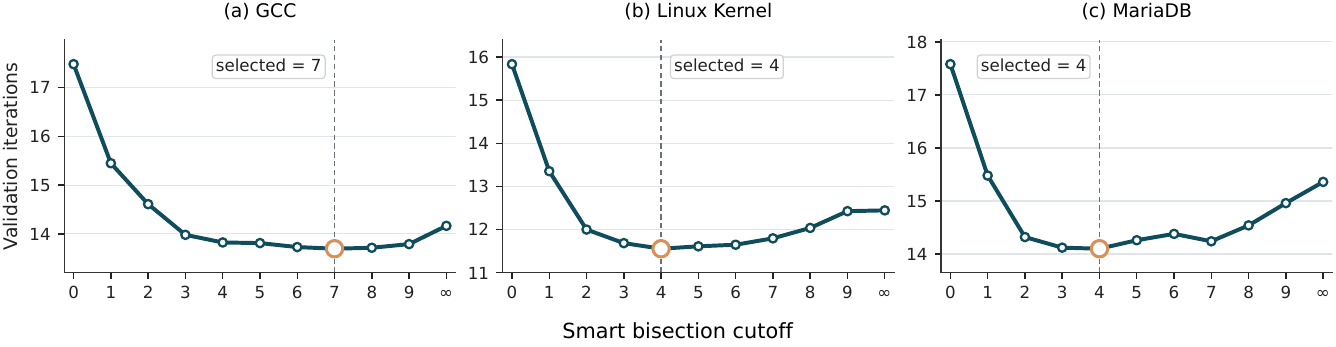}
\vspace{-3pt}

\caption{Sensitivity of \sys to the smart bisection cutoff $\tau$. Each subplot reports the average number of validation iterations under different cutoff values. The highlighted point indicates the validation-selected cutoff.}
\label{fig:sensitivity-matrix}
\end{figure*}

RQ2 examines whether \sys depends on a finely tuned smart bisection cutoff. Figure~\ref{fig:sensitivity-matrix} reports the average validation iterations for each candidate cutoff. The highlighted point in each subplot is the cutoff selected by the validation protocol in Section~\ref{sec_eval_setup}: $\tau=7$ for \emph{GCC}, $\tau=4$ for \emph{MariaDB}, and $\tau=4$ for the \emph{Linux kernel}.

The curves show that a small number of probability-driven steps captures most of the benefit. Moving from $\tau=0$ to positive cutoffs sharply reduces validation iterations across all three projects, indicating that historical latency is most useful during the early, large-interval phase of bisection. However, the best results occur at moderate cutoffs rather than at $\tau=\infty$, supporting the fallback design: after several probability-guided pivots, standard median bisection is more reliable than sparse local historical weights.

To interpret why moderate cutoffs are sufficient, we measure the median remaining interval size at the selected cutoff. At the selected cutoff, the median remaining interval is 106 commits for \emph{GCC}, 214.5 for \emph{MariaDB}, and 115 for the \emph{Linux kernel}. Thus, by the time \sys falls back to standard median bisection, the residual search has usually been reduced to only hundreds of commits.

The selected cutoffs are also not isolated optima. Validation performance remains nearly flat around $\tau=6$--$8$ for \emph{GCC}, around the selected moderate cutoff for \emph{MariaDB}, and around $\tau=4$--$6$ for the \emph{Linux kernel}. This plateau behavior indicates that \sys does not require delicate tuning; it mainly needs a moderate cutoff that uses historical latency while the interval is large and then lets median bisection finish the narrowed residual search.

\begin{takeawaybox}
\textbf{Answer to \textit{RQ2}:}
Moderate cutoffs perform best, and \sys does not require delicate $\tau$ tuning.
\end{takeawaybox}

\subsection{RQ3: Robustness to Noisy Historical Data}
\label{sec_robustness}

\sys relies on historical BIC latencies, so its effectiveness depends on whether the historical ledger remains useful when records are incomplete or duplicated. RQ3 evaluates this concern by perturbing the training history while keeping the validation and test partitions unchanged.

We study random record perturbations. For a noise level $\alpha$, we randomly select an $\alpha$ fraction of the training BIC records. Half of the selected records are removed to simulate missing historical BICs, and the other half are duplicated to simulate repeated reports. We evaluate $\alpha \in {0.00,0.10,0.25,0.50}$ and report averages over 20 random seeds.

\begin{table}[h]
    \caption{Robustness under varying levels of random data degradation (Noise level indicates the fraction of training records deleted/duplicated).}
    \label{tbl_robustness}
    \vspace{-5pt}
    \centering
    \setlength{\tabcolsep}{4pt}
    \footnotesize
    \begin{tabular}{@{}llrrr@{}}
        \toprule
        \textbf{Project} & \textbf{Noise} & \textbf{Avg. Iter.} & \textbf{Saved Iter.} & \textbf{Degrad.} \\
        \midrule
        GCC & 0.00 & 13.48 & 4.12 & 0.00 \\
        GCC & 0.10 & 13.47 & 4.13 & -0.01 \\
        GCC & 0.25 & 13.48 & 4.12 & 0.00 \\
        GCC & 0.50 & 13.48 & 4.12 & 0.00 \\
        \midrule
        MariaDB & 0.00 & 13.15 & 4.59 & 0.00 \\
        MariaDB & 0.10 & 13.21 & 4.53 & 0.06 \\
        MariaDB & 0.25 & 13.24 & 4.50 & 0.09 \\
        MariaDB & 0.50 & 13.18 & 4.56 & 0.03 \\
        \midrule
        Linux & 0.00 & 11.55 & 4.48 & 0.00 \\
        Linux & 0.10 & 11.52 & 4.51 & -0.03 \\
        Linux & 0.25 & 11.54 & 4.49 & -0.01 \\
        Linux & 0.50 & 11.53 & 4.50 & -0.02 \\
        \bottomrule
    \end{tabular}
    \vspace{-10pt}
\end{table}

Table~\ref{tbl_robustness} shows that random perturbations have little effect on \sys. Even when $\alpha=0.50$, the change in average iterations remains close to zero across all three projects. The largest observed change in Table~\ref{tbl_robustness} is below 0.1 iterations, and several perturbations even slightly improve the average iteration count. These small fluctuations indicate that \sys is not driven by individual historical records; its pivot choices are mainly determined by the coarse cumulative shape of the latency distribution.


\begin{takeawaybox}
\textbf{Answer to \textit{RQ3}:}
\sys remains stable under substantial deletion and duplication noise.
\end{takeawaybox}

\section{Threats to Validity}
\label{sec_threats}

\noindent\textbf{Internal Validity.}
The main internal threat is the accuracy of BIC extraction from unstructured bug-report discussions. We mitigate this threat with repository-level checks, including commit-existence validation, hash canonicalization, and chronological consistency checks that filter out fixing commits, post-report testing commits, and invalid hashes. We also manually inspected a subset of extracted records and found no false positives in that subset. Residual extraction noise may remain, but our robustness study shows that \sys is stable under substantial random deletion and duplication of historical records. A second internal threat is temporal leakage. We address it by using a chronological train/validation/test split: historical records initialize the prior, validation records select the smart bisection cutoff, and the newest records are reserved for held-out testing.

\smallskip
\noindent\textbf{External Validity.}
Our study uses three mature software projects: \emph{GCC}, the \emph{Linux kernel}, and \emph{MariaDB}. These projects are appropriate for studying bisection because they have long commit histories and costly historical builds. However, their BIC latency distributions may not represent all software domains. Projects with short release cycles, squash-heavy workflows, or substantially different reporting practices may exhibit different temporal patterns. Our results therefore support the existence and exploitability of temporal BIC skew in the studied systems, rather than claiming universality across all projects.

\section{Discussion}
\label{sec_discussion}

\smallskip
\noindent\textbf{Beyond Latency.}
This paper instantiates probability-driven bisection using commit latency
because it is lightweight, language-agnostic, and readily available from
version-control histories. However, the weighted-median formulation is
not tied to latency. More generally, pivot selection can be viewed as a
prior-estimation problem: commit latency provides a simple one-dimensional
prior, but future systems can incorporate richer project-specific
signals, such as code churn, subsystem activity, developer ownership,
release phases, or component-specific bug histories, into commit weights.
These signals change only the order in which commits are tested;
the final good--bad decisions are still made by the same execution
oracle, preserving the correctness guarantees inherited from standard
bisection.

\smallskip
\noindent\textbf{Beyond Bisection.}
Although this paper focuses on BIC identification, the empirical temporal skew observed in Section~\ref{sec_empirical_study} has broader impacts. If bugs are disproportionately introduced by recent changes, software testing and maintenance strategies can benefit from prioritizing recently modified components instead of treating all code regions uniformly. 
For example, new fuzzing approaches can incorporate the temporal latency of historical bugs into their mutation strategies, allocating more testing effort to execute paths through recently changed code, which has a higher probability of inducing bugs.
Exploring such applications lies beyond the scope of this paper, but our results suggest that temporal information can provide useful guidance beyond commit-history search itself.

\smallskip
\noindent\textbf{Developer-Side Data Collection.}
Our study reconstructs BIC latencies from public bug-report histories,
but deployed projects need not rely on retrospective reconstruction. In
deployment, projects can maintain structured regression records that
store the resolved BIC together with the report-time reference commit
and the searched commit sequence, from which BIC latency can be computed
directly. Such records provide a cleaner historical prior than
public discussion data, potentially strengthening the probability signal
available to \sys.

\section{Related Work}
\label{sec_related_work}
Fault localization and bisection are related but not identical debugging tasks. Fault-localization techniques typically rank suspicious code entities, executions, configurations, or changes, while bisection identifies the historical commit that first introduced a fault. We discuss this literature because several recent debugging pipelines either rely on BICs or use BIC-derived signals, making the cost of obtaining BICs relevant to their practical use. In contrast to methods that require bug reports, spectra, witnesses, configuration variation, or learned representations, \sys preserves the standard good–bad bisection interface and changes only the pivot-selection rule. Thus, \sys is best viewed as a lightweight accelerator for BIC identification that can complement, rather than replace, downstream fault-localization techniques.
\subsection{IR-Based and Learning-Based Bug Localization}

IR-based bug localization ranks suspicious code entities by matching natural-language bug reports against source-code artifacts. Classic and industrial systems exploit lexical similarity, structured code information, query reformulation, repository context, and benchmark-driven evaluation to retrieve likely faulty files or commits~\cite{zhou2012buglocator,saha2013bluir,rahman2018blizzard,wang2015evaluating,lee2018bench4bl,murali2021industryscaleirbasedbuglocalization}. More recent learning-based approaches use neural representations, relevance feedback, semantic-flow information, contrastive learning, or cross-project/cross-language models to improve bug localization when textual overlap is weak~\cite{samir2025improvedirbasedbuglocalization,du2023semanticcodebert,chakraborty2025blaze}.

These techniques narrow the debugging search space through textual or
learned similarity, but they generally return ranked candidates that
require further inspection or validation. \sys instead keeps the
execution-confirmed bisection oracle and reduces the number of commits
that must be tested.

\subsection{Execution-Based and History-Aware Fault Localization}

Spectrum-based fault localization (SBFL) ranks suspicious code entities by comparing their execution behavior across passing and failing runs~\cite{abreu2007accuracy,sasaki2020sbfl,dutta2021combi}. In compiler debugging, witness-based techniques such as \emph{DiWi}, \emph{RecBi}, and \emph{LLM4CBI} generate additional witness programs to obtain contrasting executions~\cite{chen2019diwi,chen2020recbi,tu2024llm4cbi}. Configuration-driven techniques such as \emph{ODFL} instead vary compiler configurations to collect differentiating spectra~\cite{yang2025odfl}. \emph{Fonte} is also closely related because it targets BIC identification from faults: it combines fault-localization signals with code-change relationships to rank candidate inducing commits~\cite{an2023fonte}.

Other work incorporates historical information into fault localization. \emph{HSFL} uses historical bug-inducing and non-inducing commits to adjust suspiciousness scores~\cite{wen2019hsfl}. \emph{BASIC} uses files modified by an identified BIC as fault candidates, showing that BICs can provide effective developer-aligned localization signals~\cite{yang2025using}. \emph{DualTrack} further combines BIC-derived historical information with SBFL-style dynamic spectra, motivated by the observation that these two sources provide complementary localization signals~\cite{li2026dualtrack}.

These approaches provide strong localization signals, but they require
artifacts beyond standard bisection, such as spectra, witnesses,
configuration contrasts, or already identified BICs. \sys is
complementary: it reduces the cost of obtaining the inducing commit
itself using the same good--bad oracle as conventional bisection.

\subsection{Changeset-Based Bug Localization}

Changeset-based bug localization aims to identify the change responsible for a bug report, typically by ranking commits or changesets according to their similarity to the report. Recent approaches use learned change representations, BERT-style encoders, semantic code graphs, data augmentation, online adaptation, or learning-to-rank models to match bug reports with likely inducing changes~\cite{ciborowska2022fast,ciborowska2023bugreportsexploringdata,ciborowska2022onlineadaptablebuglocalization,du2023semanticcodebert,wen2016locus,loyola2018rankchanges}. These techniques are close to \sys in granularity because they reason about commits or changesets rather than only files or statements.

These techniques are related to \sys because both exploit version-history
information when reasoning about faults. However, their objective and
interface differ. Changeset-based localization predicts or ranks source
changes that are likely to be relevant to a bug report, and the returned
candidates still require developer inspection or execution-based
validation. \sys does not rank changes by textual or learned relevance.
It accelerates bisection by changing which historical commit
is tested next, while each good–bad decision and the final BIC remain
determined by the bug-triggering test.

\section{Conclusion}
\label{sec_conclusion}

This paper revisits an assumption behind standard bisection: balancing commit count is a good proxy for balancing BIC likelihood. By analyzing 8,172 developer-reported BICs from \emph{GCC}, the \emph{Linux kernel}, and \emph{MariaDB}, we found that BICs are strongly concentrated near recent history, with the median BIC appearing within the most recent 0.69\% of commits. Motivated by this temporal skew, we introduced \sys, a probability-driven bisection approach that uses historical BIC latency to select weighted-median pivots. In the evaluation, \sys reduces bisection iterations by 25.75\% on average across projects and improves over standard bisection in 91.26\% of test cases. 
We expect that \sys can significantly save debugging efforts of developers in practice, and more broadly, the insights about the BIC distribution behind this paper can benefit future software engineering research.


\balance
\bibliographystyle{IEEEtran}
\bibliography{ref}

@manual{git-bisect-manual,
  title        = {git-bisect — Use binary search to find the commit that introduced a bug},
  author       = {{The Git Development Community}},
  organization = {git-scm.com},
  year         = {2025},
  note         = {Online; accessed 2025-09-10},
  url          = {https://git-scm.com/docs/git-bisect}
}

@article{courtiel2024theoretical,
  title={Theoretical analysis of git bisect},
  author={Courtiel, Julien and Dorbec, Paul and Lecoq, Romain},
  journal={Algorithmica},
  volume={86},
  number={5},
  pages={1365--1399},
  year={2024},
  publisher={Springer}
}

@inproceedings{an2023fonte,
  title={Fonte: Finding bug inducing commits from failures},
  author={An, Gabin and Hong, Jingun and Kim, Naryeong and Yoo, Shin},
  booktitle={2023 IEEE/ACM 45th International Conference on Software Engineering (ICSE)},
  pages={589--601},
  year={2023},
  organization={IEEE}
}

@article{sliwerski2005changes,
  title={When do changes induce fixes?},
  author={{\'S}liwerski, Jacek and Zimmermann, Thomas and Zeller, Andreas},
  journal={ACM sigsoft software engineering notes},
  volume={30},
  number={4},
  pages={1--5},
  year={2005},
  publisher={ACM New York, NY, USA}
}

@inproceedings{ciborowska2022fast,
  title={Fast changeset-based bug localization with BERT},
  author={Ciborowska, Agnieszka and Damevski, Kostadin},
  booktitle={Proceedings of the 44th international conference on software engineering},
  pages={946--957},
  year={2022}
}

@article{yang2025using,
  title={Using a Sledgehammer to Crack a Nut? Revisiting Automated Compiler Fault Isolation},
  author={Yang, Yibiao and Li, Qingyang and Sun, Maolin and Wu, Jiangchang and Zhou, Yuming},
  journal={arXiv preprint arXiv:2512.16335},
  year={2025}
}

@inproceedings{murali2021industryscaleirbasedbuglocalization,
author = {Murali, Vijayaraghavan and Gross, Lee and Qian, Rebecca and Chandra, Satish},
title = {Industry-scale IR-based bug localization: a perspective from Facebook},
year = {2021},
isbn = {9780738146690},
publisher = {IEEE Press},
url = {https://doi.org/10.1109/ICSE-SEIP52600.2021.00028},
doi = {10.1109/ICSE-SEIP52600.2021.00028},
booktitle = {Proceedings of the 43rd International Conference on Software Engineering: Software Engineering in Practice},
pages = {188–197},
numpages = {10},
location = {Virtual Event, Spain},
series = {ICSE-SEIP '21}
}

@misc{samir2025improvedirbasedbuglocalization,
      title={Improved IR-based Bug Localization with Intelligent Relevance Feedback}, 
      author={Asif Mohammed Samir and Mohammad Masudur Rahman},
      year={2025},
      eprint={2501.10542},
      archivePrefix={arXiv},
      primaryClass={cs.SE},
      url={https://arxiv.org/abs/2501.10542}, 
}

@misc{ciborowska2022onlineadaptablebuglocalization,
      title={Online Adaptable Bug Localization for Rapidly Evolving Software}, 
      author={Agnieszka Ciborowska and Michael J. Decker and Kostadin Damevski},
      year={2022},
      eprint={2203.03544},
      archivePrefix={arXiv},
      primaryClass={cs.SE},
      url={https://arxiv.org/abs/2203.03544}, 
}

@inproceedings{du2023semanticcodebert,
author = {Du, Yali and Yu, Zhongxing},
title = {Pre-training Code Representation with Semantic Flow Graph for Effective Bug Localization},
year = {2023},
isbn = {9798400703270},
publisher = {Association for Computing Machinery},
address = {New York, NY, USA},
url = {https://doi.org/10.1145/3611643.3616338},
doi = {10.1145/3611643.3616338},
booktitle = {Proceedings of the 31st ACM Joint European Software Engineering Conference and Symposium on the Foundations of Software Engineering},
pages = {579–591},
numpages = {13},
location = {San Francisco, CA, USA},
series = {ESEC/FSE 2023}
}

@misc{ciborowska2023bugreportsexploringdata,
    title={Too Few Bug Reports? Exploring Data Augmentation for Improved Changeset-based Bug Localization}, 
      author={Agnieszka Ciborowska and Kostadin Damevski},
      year={2023},
      eprint={2305.16430},
      archivePrefix={arXiv},
      primaryClass={cs.SE},
      url={https://arxiv.org/abs/2305.16430}, 
}

@inproceedings{wen2020boosting,
  title={Boosting automated program repair with bug-inducing commits},
  author={Wen, Ming and Liu, Yepang and Cheung, Shing-Chi},
  booktitle={Proceedings of the ACM/IEEE 42nd International Conference on Software Engineering: New Ideas and Emerging Results},
  pages={77--80},
  year={2020}
}

@INPROCEEDINGS{7816476,
  author={Shimagaki, Junji and Kamei, Yasutaka and McIntosh, Shane and Pursehouse, David and Ubayashi, Naoyasu},
  booktitle={2016 IEEE International Conference on Software Maintenance and Evolution (ICSME)}, 
  title={Why are Commits Being Reverted?: A Comparative Study of Industrial and Open Source Projects}, 
  year={2016},
  volume={},
  number={},
  pages={301-311},
  doi={10.1109/ICSME.2016.83}}

@article{dutta2021combi,
  title={Combi-FL: Neural network and SBFL based fault localization using mutation analysis},
  author={Dutta, Arpita and Srivastava, Saksham Sahai and Godboley, Sangharatna and Mohapatra, Durga Prasad},
  journal={Journal of Computer Languages},
  volume={66},
  pages={101064},
  year={2021},
  publisher={Elsevier}
}

@inproceedings{sasaki2020sbfl,
  title={Sbfl-suitability: A software characteristic for fault localization},
  author={Sasaki, Yui and Higo, Yoshiki and Matsumoto, Shinsuke and Kusumoto, Shinji},
  booktitle={2020 IEEE International Conference on Software Maintenance and Evolution (ICSME)},
  pages={702--706},
  year={2020},
  organization={IEEE}
}

@inproceedings{wang2015evaluating,
  title={Evaluating the usefulness of ir-based fault localization techniques},
  author={Wang, Qianqian and Parnin, Chris and Orso, Alessandro},
  booktitle={Proceedings of the 2015 international symposium on software testing and analysis},
  pages={1--11},
  year={2015}
}

@inproceedings{abreu2007accuracy,
  title={On the accuracy of spectrum-based fault localization},
  author={Abreu, Rui and Zoeteweij, Peter and Van Gemund, Arjan JC},
  booktitle={Testing: Academic and industrial conference practice and research techniques-MUTATION (TAICPART-MUTATION 2007)},
  pages={89--98},
  year={2007},
  organization={IEEE}
}

@inproceedings{lee2018bench4bl,
  title={Bench4bl: reproducibility study on the performance of ir-based bug localization},
  author={Lee, Jaekwon and Kim, Dongsun and Bissyand{\'e}, Tegawend{\'e} F and Jung, Woosung and Le Traon, Yves},
  booktitle={Proceedings of the 27th ACM SIGSOFT international symposium on software testing and analysis},
  pages={61--72},
  year={2018}
}

@article{wen2019hsfl,
  title={Historical Spectrum Based Fault Localization},
  author={Wen, Ming and Chen, Junjie and Tian, Yongqiang and Wu, Rongxin and Hao, Dan and Han, Shi and Cheung, Shing-Chi},
  journal={IEEE Transactions on Software Engineering},
  volume={47},
  number={11},
  pages={2348--2368},
  year={2021},
  publisher={IEEE}
}

@article{li2026dualtrack,
  title={Boosting Compiler Fault Localization: Getting the Best of Both Worlds by Fusing Dynamic and Historical Data},
  author={Li, Qingyang and Yang, Yibiao and Sun, Maolin and Wu, Jiangchang and Shi, Qingkai and Zhou, Yuming and Xu, Baowen},
  journal={IEEE Transactions on Software Engineering},
  volume={52},
  number={4},
  pages={1462--1477},
  year={2026},
  publisher={IEEE}
}

@article{yang2025odfl,
  title={Isolating Compiler Faults Through Differentiated Compilation Configurations},
  author={Yang, Yibiao and Li, Qingyang and Sun, Maolin and Yang, Jing and Wu, Jiangchang and Zhou, Yuming},
  journal={IEEE Transactions on Software Engineering},
  volume={51},
  number={6},
  pages={1838--1853},
  year={2025},
  publisher={IEEE},
  doi={10.1109/TSE.2025.3569530}
}

@inproceedings{chen2019diwi,
  title={Compiler Bug Isolation via Effective Witness Test Program Generation},
  author={Chen, Junjie and Han, Jiaqi and Sun, Peiyi and Zhang, Lingming and Hao, Dan and Zhang, Lu},
  booktitle={Proceedings of the 2019 27th ACM Joint Meeting on European Software Engineering Conference and Symposium on the Foundations of Software Engineering},
  series={ESEC/FSE 2019},
  pages={223--234},
  year={2019},
  publisher={Association for Computing Machinery},
  doi={10.1145/3338906.3338957}
}

@inproceedings{chen2020recbi,
  title={Enhanced Compiler Bug Isolation via Memoized Search},
  author={Chen, Junjie and Ma, Haoyang and Zhang, Lingming},
  booktitle={Proceedings of the 35th IEEE/ACM International Conference on Automated Software Engineering},
  series={ASE 2020},
  pages={78--89},
  year={2020},
  publisher={Association for Computing Machinery},
  doi={10.1145/3324884.3416570}
}

@article{tu2024llm4cbi,
  title={Isolating Compiler Bugs by Generating Effective Witness Programs with Large Language Models},
  author={Tu, Haoxin and Zhou, Zhide and Jiang, He and Yusuf, Imam Nur Bani and Li, Yuxian and Jiang, Lingxiao},
  journal={IEEE Transactions on Software Engineering},
  volume={50},
  number={7},
  pages={1768--1788},
  year={2024},
  publisher={IEEE},
  doi={10.1109/TSE.2024.3403958}
}

@inproceedings{saha2017selective,
  title={Selective Bisection Debugging},
  author={Saha, Ripon and Gligoric, Milos},
  booktitle={Fundamental Approaches to Software Engineering},
  pages={60--77},
  year={2017},
  publisher={Springer}
}

@inproceedings{zhou2012buglocator,
  title     = {Where Should the Bugs Be Fixed? More Accurate Information Retrieval-Based Bug Localization Based on Bug Reports},
  author    = {Zhou, Jian and Zhang, Hongyu and Lo, David},
  booktitle = {Proceedings of the 34th International Conference on Software Engineering},
  series    = {ICSE 2012},
  pages     = {14--24},
  year      = {2012},
  publisher = {IEEE Press}
}

@inproceedings{saha2013bluir,
  title     = {Improving Bug Localization Using Structured Information Retrieval},
  author    = {Saha, Ripon K. and Lease, Matthew and Khurshid, Sarfraz and Perry, Dewayne E.},
  booktitle = {Proceedings of the 28th IEEE/ACM International Conference on Automated Software Engineering},
  series    = {ASE 2013},
  pages     = {345--355},
  year      = {2013},
  publisher = {IEEE},
  doi       = {10.1109/ASE.2013.6693093}
}

@inproceedings{rahman2018blizzard,
  title     = {Improving IR-Based Bug Localization with Context-Aware Query Reformulation},
  author    = {Rahman, Mohammad Masudur and Roy, Chanchal K.},
  booktitle = {Proceedings of the 2018 26th ACM Joint Meeting on European Software Engineering Conference and Symposium on the Foundations of Software Engineering},
  series    = {ESEC/FSE 2018},
  pages     = {621--632},
  year      = {2018},
  publisher = {Association for Computing Machinery},
  doi       = {10.1145/3236024.3236065}
}

@inproceedings{wen2016locus,
  title     = {Locus: Locating Bugs from Software Changes},
  author    = {Wen, Ming and Wu, Rongxin and Cheung, Shing-Chi},
  booktitle = {Proceedings of the 31st IEEE/ACM International Conference on Automated Software Engineering},
  series    = {ASE 2016},
  pages     = {262--273},
  year      = {2016},
  publisher = {Association for Computing Machinery},
  doi       = {10.1145/2970276.2970359}
}

@inproceedings{loyola2018rankchanges,
  title     = {Bug Localization by Learning to Rank and Represent Bug-Inducing Changes},
  author    = {Loyola, Pablo and Gajananan, Kugamoorthy and Satoh, Fumiko},
  booktitle = {Proceedings of the 27th ACM International Conference on Information and Knowledge Management},
  series    = {CIKM 2018},
  pages     = {657--665},
  year      = {2018},
  publisher = {Association for Computing Machinery},
  doi       = {10.1145/3269206.3271811}
}

@ARTICLE{chakraborty2025blaze,
  author={Chakraborty, Partha and Alfadel, Mahmoud and Nagappan, Meiyappan},
  journal={IEEE Transactions on Software Engineering}, 
  title={BLAZE: Cross-Language and Cross-Project Bug Localization via Dynamic Chunking and Hard Example Learning}, 
  year={2025},
  volume={51},
  number={8},
  pages={2254-2267},
  doi={10.1109/TSE.2025.3579574}}

\end{document}